\documentclass[10pt,aps,prl,amssymb,twocolumn,floatfix,superscriptaddress]{revtex4-1}
\usepackage{amsmath,bm,epsfig,graphicx}
\usepackage{booktabs,multirow}
\usepackage[colorlinks,linkcolor=blue,citecolor=blue]{hyperref} 

\newcommand{\noi}{\noindent}

\begin{document}
\title{Observation of Electrically Tunable van der Waals Interaction in Graphene-Molecule Complex}
\author{Manoharan Muruganathan}
\thanks{mano@jaist.ac.jp}
\affiliation{School of Materials Science, Japan Advanced Institute of
Science and Technology, Nomi 923-1211, Japan}
\author{Jian Sun}
\affiliation{School of Materials Science, Japan Advanced Institute of
Science and Technology, Nomi 923-1211, Japan}
\author{Tomonori Imamura}
\affiliation{School of Materials Science, Japan Advanced Institute of
Science and Technology, Nomi 923-1211, Japan}
\author{Hiroshi Mizuta}
\affiliation{School of Materials Science, Japan Advanced Institute of
Science and Technology, Nomi 923-1211, Japan}
\affiliation{Nanoelectronics and Nanotechnologies Research Group, Faculty of Physical Sciences and Engineering, University of 
Southampton, Highfield, Southampton SO17 1BJ, UK}
\date{\today}

\begin{abstract}
van der Waals (vdW) interaction plays a fundamental role in the surface-molecules related phenomena. Tuning of the correlated charge fluctuation in the vdW complex is a plausible way to modulate the molecules interaction at the atomic surface. We report vdW interaction tunability of the graphene-CO$_2$ complex by combining the first principle calculations with the vdW exchange correlation density functionals and the time evaluation measurements of CO$_2$ molecules adsorption/desorption on graphene under an external electric field. The field-dependent charge transfer within the complex unveils the controllable tuning of CO$_2$ from acceptor to donor. Meanwhile the configuration of the adsorbed molecule - the equilibrium distance from graphene and O-C-O bonding angle - is modified accordingly. The range of electrical tunability is a unique feature for each type of molecules.
\end{abstract}
\maketitle

van der Waals (vdW) interaction plays an important role in the surface-related physics and chemistry \cite{LIFSHITZ-Sov-phy-usp-1961}. It is the decisive factor in the molecular physisorption on a surface \cite{Nat.Chem.-2010-bartels, Aradhya-Nat.Mat.-2012}, adhesion between micro-machined surfaces in the micro- and nano-mechanical devices \cite{naturematerials-Delrio-2005}, friction in tribology discipline \cite{AP-Krim-2012}, and the characteristics in the vdW heterostructures \cite{Nat.-Geim-2013, Science-Dai-2014}. At present, surface modification, i.e. tailoring or replacing the original materials, is a commonly employed method to selectively create specific vdW complexes for their favorable properties \cite{Nat.Mat.-2004-LALATONNE, Lessel-PRL-2013}.  

The London dispersion forces involved in a vdW complex of molecules adsorbed on a surface is electro-dynamically correlated to the charge transfer inside the complex. This raises the possibility of tuning such vdW interaction electrically, e.g. by applying external electrical field, which offers an easy access to the desired features while keeping the original materials untouched\cite{Chem.Rev.-Dethlefs-2000, PNAS-Zhou-2010, Phys.Chem.Chem.Phys-Nejad-2013}. However, we are not aware of any experimental reports on vdW interaction showing the electrical tunability. The difficulty of characterizing the vdW complex impedes the direct observation. In this report, we demonstrate the electrical tunability of the vdW interaction for graphene-molecule complexes with first principle calculations combined with the experimental study of the electrically controlled charge transfer for the graphene-CO$_2$ vdW complex.
As a low-noise Dirac fermion material, graphene is strongly modulated by the vdW interactions with physisorbed molecules upon its transport properties; the inaccessible interaction is therefore transferred into the easily measurable electric signal \cite{Novoselov-Science-2004, Schedin-Nat.Mats.-2007}. The transport measurements verify the tunability of vdW interaction between graphene and CO$_2$ molecules from the varied charge transfer from molecules to graphene under external fields.

\begin{figure}[t]
\centering
\includegraphics[scale=0.45]{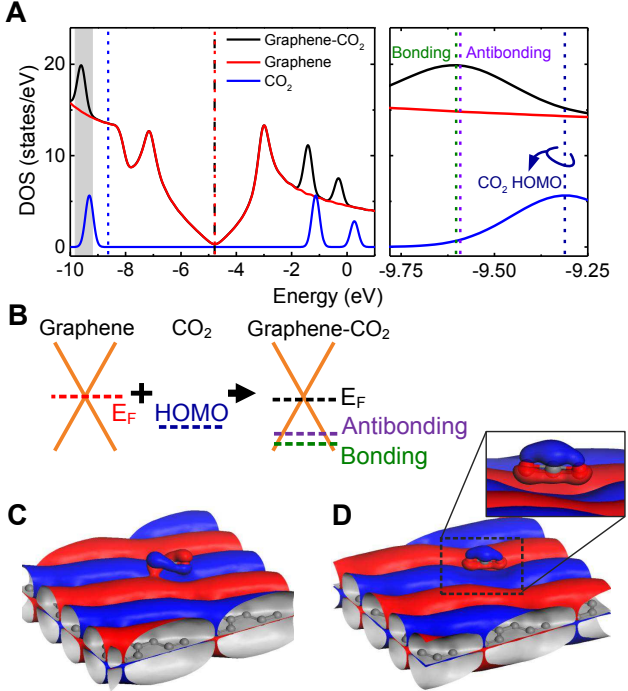}
\caption{Density of States and wavefunctions of a graphene-CO$_2$ complex. (A) DOS plots of graphene-CO$_2$ complex, graphene, and CO$_2$ molecule. Vertical dash-lines locate the their Fermi levels at -4.7840 eV, -4.7788 eV, and -8.6309 eV, respectively. The right panel is the zoomed-in details of the grey region. (Gaussian broadening energy in DOS plots: 0.2 eV) (B) Schematic of the bonding-antibonding splitting in graphene-CO$_2$ complex. (C) and (D) visualize bonding (-9.60327 eV) and anti-Bonding (-9.59135 eV) Gamma point wavefunctions of graphene-CO$_2$ complex, respectively. Blue and red colors denote the positive and negative iso-surface (Iso-value: 0.005).}
\label{fig:fig1}
\end{figure}

First, the first principle calculations are utilized to understand the interaction between graphene and a CO$_2$ molecule adsorbed on it \cite{Soler-J.Phys.Condens.Matter-2002}. The nonlocal dispersion relationship associated with the molecular adsorption on graphene is described by using van der Waals density functionals reported by Dion \cite{Dion-PRL-2004, Roman-Perez-PRL-2009}. The potential energy surface scans show the adsorbed CO$_2$ molecule lie most stably in the direction perpendicular to the carbon-carbon bond of the graphene hexagon, sitting at an equilibrium distance of 3.2 \AA~ with an energy of 170 meV. The low adsorption energy reflects a weak interaction between the CO$_2$ molecule and the graphene and its physisorption nature. This is also revealed by the density of states (DOS) calculations \cite{Bogani-Nat.Mat.-2008, Wiley-Masel, Zhang-Nanotech-2009} as the electronic states of CO$_2$ are superimposed on to these of the graphene (Fig. \ref{fig:fig1}A). The weak interaction of the highest occupied molecular orbital (HOMO) of CO$_2$ (the peak showed at -9.31159 eV in Fig. \ref{fig:fig1}A) with the graphene states results in a small bonding-antibonding splitting of 11.92 meV (Fig. \ref{fig:fig1}B). The 3-dimensional plots of the wavefuntions visually elucidate these abstract features. Figure \ref{fig:fig1}C depicts the bonding wavefunction at -9.60327 eV showing a weak graphene-CO$_2$ orbitals overlap while a repulsion is noted in its antibonding wavefunction at -9.59135 eV as showed in Fig. \ref{fig:fig1}D. Furthermore, these interacting CO$_2$ states still locate energetically far below the Fermi level clarifies the weak charge transfer within the graphene-CO$_2$ complex. Consequently, the van der Waals interaction plays a dominant factor in the graphene-CO$_2$ complex interaction

\begin{figure}[t]
\centering
\includegraphics[scale=0.6]{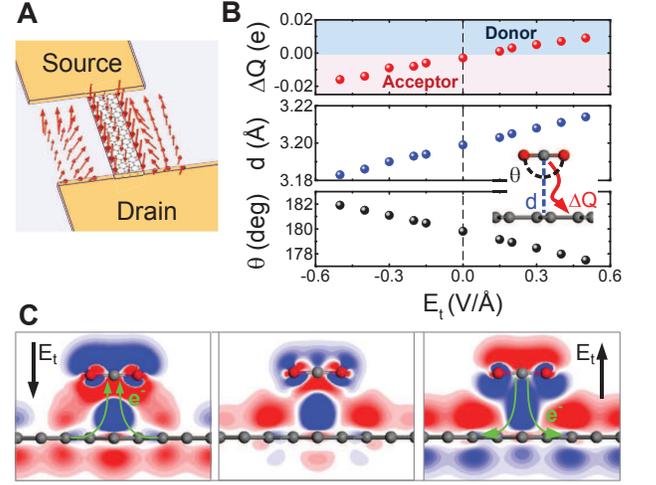}
\caption{Electrical tuning of van der Waals bonding of CO$_2$ molecule to graphene. (A) Finite element simulation showing the direction of electric filed at the surface of 2 \AA ~above graphene with 40 V applied back gate voltage. (B) First principle calculations of charge transfer $\Delta Q$, bonding distance $d$ and bending angle $\theta$ of the adsorbed CO$_2$ molecule on graphene at various tuning electric fields. Inset illustrates a absorbed molecule. (C) Charge density distributions in the cut planes across the CO$_2$ at varied electrical fields (tuning voltages). From left to right: -0.15 V/\AA~(40 V), 0 V/\AA~(0 V), and 0.15 V/\AA~(-40 V). Blue color denotes electron density enrichment and red color shows depletion for the iso-values of $\pm$ 0.0001 electrons/\AA$^{3}$.}
\label{fig:fig2}
\end{figure}

Now, we unveil the electrical tunability of the graphene-CO$_2$ molecule vdW complex by carrying out first principles calculation with Mulliken population analysis \cite{Kristian-PRB-2013}. The vdW complex is investigated under external electric field, $E_t$, applied perpendicularly to graphene.  
In practice, the electric field is usually introduced to graphene surface by applying voltage, $V_T$ to the substrate. A finite element simulation is first employed to estimate the strength of electric field generated with the application of $V_T$, which shows $E_t$ of -0.15 V/\AA~(0.15 V/\AA)~is achieved on the graphene surface for $V_T$ of 40 V (-40 V) (Fig. \ref{fig:fig2}A). The charge transfer from the CO$_2$ molecule to graphene, $\Delta Q$, is calculated with varying $E_t$. Figure \ref{fig:fig2}B plots the calculated charge transfer as a function of electric field. Without applying a tuning field, CO$_2$ molecule acts as a weak acceptor to graphene, receiving 0.003$e$ charge. The varied $\Delta Q$ is noted as $E_t$ is introduced, disclosing the nature of tunable charge transfer in the vdW complex. In particular, the CO$_2$ molecules switch their role to donor in the field of $0.15$ V/\AA~(Fig. \ref{fig:fig2}B). The switched polarity of the graphene-CO$_2$ molecule complex system under the electric fields is also exhibited as a change in the electric dipole moments (figs. S10 and S11).   

\begin{figure}[t]
\centering
\includegraphics[scale=0.52]{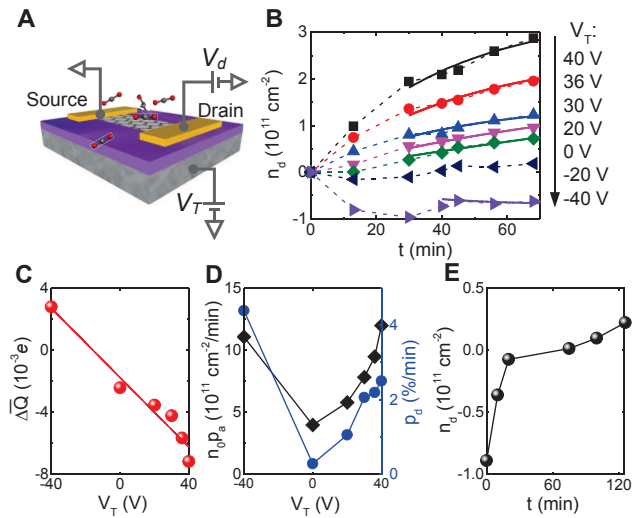}
\caption{Measurements of CO$_2$ adsorptions at varied tuning voltages $V_T$. (A) Schematic of the measurement configuration. Tuning voltage, $V_T$, is applied at substrate. (B) Doping concentration at various tuning voltages, $V_T$, from 40 V to -40 V. The positive/negative sign of $n_d$ denotes hole/electron. The solid lines are the fits to Eq.(\ref{eq:ad-de}). 
(C) and (D) plot the values of averaged change transfer $\overline{\Delta Q}$, adsorption rate $n_0p_a$ and desorption rate $p_d$, respectively. Solid line in (C) is the linear fit.
(E) Change of doping concentration after tuning voltage switched from -40 V to 40 V.}
 \label{fig:fig3}
\end{figure}

The predicted tunable charge transfer in the vdW complex, if any, should be measured as varied doping concentration in graphene under the tuning electric fields.
In order to prove the above theoretical findings, we measure the variations of doping concentrations, $n_d$, in graphene exposed to CO$_2$ molecules under various values of $V_T$. Figure \ref{fig:fig3}A illustrates the schematic of the device and measurement configuration. The gate modulation measurements are carried out along with the time, from which the doping concentrations are extracted (Fig. \ref{fig:fig3}B). Thermal annealing at $\sim 10^{-4}$ Pa is employed \textit{in situ} to regenerate graphene by cleansing the adsorbates before each measurement \cite{Schedin-Nat.Mats.-2007}. 
For $V_T$ ranging from 0 V to 40 V, positive increases in the doping concentration $n_d$ are observed, which indicate the adsorbed CO$_2$ molecules act as acceptor. 
Interestingly, for $V_T$ of -40 V, the negative $n_d$ is measured, signifying the CO$_2$ molecules act as donor. These results strongly substantiate the prediction of the first principle calculations. 

We can extract the value of the average charge transfer $\overline{\Delta Q}$ from Fig. \ref{fig:fig3}B with a simple dynamic model \cite{Chen-NewJ.Phys-2010}.
Then, doping concentration is expressed as $n_d(t)=N(t)\overline{\Delta Q}$, where $N(t)$ represents the temporal density of CO$_2$ molecules adsorbed on graphene and is described 
\begin{align}
N(t)=\frac{n_0p_a}{p_d}(1-e^{-p_dt})+N_0,
\label{eq:ad-de}
\end{align}
where $n_0$ is the density of the molecules in the gas phase (remains constant after gas injection is stopped), $N_0$ is the initial adsorption density on graphene, $p_a$ and $p_d$ are the adsorption and desorption rates, respectively. By fitting Eq. (1) to the measured temporal change in the adsorption density, the values of $\overline{\Delta Q}$ are obtained for different $V_T$. Fig. \ref{fig:fig3}C clearly manifests the charge transfer tuned via applied external fields, showing a good agreement with the first principles calculation prediction. Besides, the adsorption and desorption rates are also extracted from the data fittings (Fig. \ref{fig:fig3}D). The adsorption rate enhanced by the increased magnitude $V_T$, accelerates the adsorption process due to the polarizability of the CO$_2$ molecules. Hence, the similar adsorption rates are read for $V_T$ of -40 V and 40 V. In contrast, the desorption rate at $V_T$ of -40 V is approximately twice as high as compared with that for $V_T$ of 40 V, as a result of a weaker vdW interaction between CO$_2$ molecule and graphene at -40 V, reflecting a longer bonding distance and lower charge transfer (Fig. \ref{fig:fig2}B). 

It is noted that the application of constant voltage $V_T$ is shortly interrupted for an interval of $\sim 0.5$ min at each individual data point in a series of temporal measurement. During the interval, $V_T$ is swept from -40 V to 40 V in order to obtain gate modulation curves for $n_d$ extraction. Under these circumstances, the charge transfer from those previously adsorbed molecules is probably altered; and molecules freshly adsorbed during the sweeping exhibit varied charge transfer different from that adsorbed under constant $V_T$. 
Here we carry out a measurement at an extreme condition to clarify the interruption of constant $V_T$ only slightly affecting the analysis of $n_d$ at constant $V_T$. Graphene is firstly kept at $V_T = -40$ V for 2 hours; a $n_d$ of $-0.89 \times 10^{11}$ cm$^{-2}$ is measured (Fig. \ref{fig:fig3}E). Then $V_T$ is suddenly switched to 40 V; the positive shift of $n_d$ with a relatively slow rate of $-0.4 \times 10^{10}$ cm$^{-2}$/min is observed. Hence, the short interruption of constant $V_T$ has only insignificant impact on the extraction of $n_d$ for $V_T$. It also suggests that the electrical tuning plays a part during the formation of physisorption or the vdW complex; that is to say once the complex formed, it is stable to some extent. 

In order to clarify the tunable charge transfer, we visualize the two-dimensional charge density difference distributions in the graphene-CO$_2$ complex in Fig. \ref{fig:fig2}C \cite{Clark-Kristallogr.-2005} under the tuning field of -0.15 V/\AA~ (left), zero (center), and 0.15 V/\AA)~(right). Here, the charge density difference of CO$_2$ adsorption is defined by $\rho_{Ad}=\rho_{CO_{2}+Gra}- \rho_{CO_{2}}-\rho_{Gra}$, where $\rho_{CO_{2}+Gra}$, $\rho_{Gra}$, and $\rho_{CO_{2}}$ are the charge densities of the graphene with adsorbed CO$_2$, the pristine graphene and an isolated CO$_2$ molecule, respectively. 
The tunability of the charge transfer can be understood by applying Lorentz force. 
For a zero tuning field, the vdW interaction governs the charge transfer between CO$_2$ and graphene, which can be seen from the induced dipole structure between them. As the tuning field is switched on, a strong electrostatic force induces a comparable order of charge redistribution inside the vdW complex; that is to say, the charge transfer is tuned electrically. Under positive (negative) tuning fields, more electrons are taken from (given to) the CO$_2$ molecule (movie S1). The reversed dipole distribution are clearly seen for the negative and positive tuning fields (left and right of Fig. \ref{fig:fig2}C). This is associated with the change in other parameters for the graphene-CO$_2$ vdW interactions. Particularly, the vdW bonding distance, $d$, is changed from 3.199 \AA~at the natural state to 3.203 \AA~(3.194 \AA) for $E_T$ of 0.15 V/\AA~(-0.15 V/\AA). The bonding angle between carbon and oxygen atoms also varies accordingly (Fig. \ref{fig:fig2}B). 

\begin{table}[t]
\centering
\caption{\label{tab:table1}Electrical tunability of the Charge transfer in various graphene-molecule complexes. Negative and Positive signs indicate the role of molecules as acceptor and donor, respectively.}
\vspace{8pt}
\begin{tabular}{cccc}
\hline
\multirow{2}{*}{Molecule} &~ & $\Delta Q$ (e) & ~ \\  \cmidrule(r){2-4}
& -0.15 V/\AA &0 V/\AA & 0.15 V/\AA \\ 
\hline
CO$_2$ &-0.006 &-0.003 &+0.001\\ 
C$_6$H$_6$ & 0.009  &+0.019  &+0.029 \\
CO & -0.003  & +0.001  & +0.004  \\
NH$_3$& +0.015  & +0.018  & +0.022  \\
O$_2$& -0.242  & -0.208 & -0.174  \\
\hline
\end{tabular}
\end{table}

We extend the present theoretical investigation to other graphene-molecules complexes. In Table.\ref{tab:table1}, we compare the their charge transfer $\Delta Q$ calculated for various vdW complexes at the negative, zero, and positive tuning fields \cite{Chakarova-Kack-PRL-2006, Schedin-Nat.Mats.-2007, Sato-Nanolett-2011}. The results indicate that the individual complexes show different tunabilities. It should also be noted that such a unique tunability does not depend on the initial conditions of graphene. Thus, additional to the conductivity change of graphene stemmed from the molecular adsorptions - a commonly employed gas sensing mechanism, the observation of the doping tunability may provide a new approach to the gas molecule detection. 

With the complex of graphene and its phsisorbed CO$_2$ molecules, we have discovered the tunability of the vdW interaction at external electric fields. The charge transfer in a vdW complex is tuned electrically, thereby changing other parameters of interaction thereupon. Our results entice us to
attempt the possibility of recognizing the type of molecules with graphene from the unique tunable change transfer between them.\\

\noi{\textbf{Acknowledgments} This work was supported by Grant-in-Aid for Scientific Research (No. 25220904) from Japan Society for the Promotion of Science and the Center of Innovation Program from Japan Science and Technology Agency. The authors thank M. E. Schmidt for the helpful discussions.}

\clearpage

\end{document}